\pdfoutput=1
\documentclass[11pt]{article}
\usepackage[utf8]{inputenc}\usepackage[T1]{fontenc}\usepackage{lmodern}
\usepackage[margin=1in]{geometry}\usepackage{amsmath,amssymb}\usepackage{graphicx}
\usepackage{booktabs}\usepackage{longtable}\usepackage{array}\usepackage{calc}\usepackage{xcolor}
\usepackage[numbers,sort&compress]{natbib}
\usepackage[colorlinks=true,linkcolor=blue,citecolor=blue,urlcolor=blue]{hyperref}
\usepackage{cleveref}
\usepackage{fancyvrb}
\usepackage{float}

\usepackage{newunicodechar}
\newunicodechar{≥}{\ensuremath{\geq}}
\newunicodechar{·}{\ensuremath{\cdot}}
\newunicodechar{⁻}{\textsuperscript{-}}
\newunicodechar{⟩}{\ensuremath{\rangle}}
\newunicodechar{⟨}{\ensuremath{\langle}}
\newunicodechar{ψ}{\ensuremath{\psi}}
\newunicodechar{→}{\ensuremath{\rightarrow}}
\newunicodechar{↳}{\ensuremath{\hookrightarrow}}
\newunicodechar{θ}{\ensuremath{\theta}}
\newunicodechar{×}{\ensuremath{\times}}
\newunicodechar{²}{\textsuperscript{2}}
\newunicodechar{¹}{\textsuperscript{1}}
\newunicodechar{⁰}{\textsuperscript{0}}
\newunicodechar{⁴}{\textsuperscript{4}}
\newunicodechar{⁵}{\textsuperscript{5}}

\providecommand{\tightlist}{\setlength{\itemsep}{0pt}\setlength{\parskip}{0pt}}

\title{When Does Fine-Tuning Beat Prompting for Qiskit-to-CUDA-Q Translation?\\ An Execution-Verified Study}
\author{%
  Ana Paula Appel\\
  Red Hat\\
  \texttt{aappel@redhat.com}
}
\date{November 2026}
\begin{document}\maketitle
\begin{abstract}
Moving quantum code from Qiskit to NVIDIA CUDA-Q is a practical adoption cost that language models are
starting to automate. We ask whether a small open model can learn this translation, and how to measure it
so that the result can be trusted. Both programs can be executed and their state vectors compared, so a
kernel counts as correct only if it prepares the same state as the Qiskit reference (fidelity
$\geq 0.999$), with no text matching and no LLM judge. We build a 640-task benchmark with three test
sets (larger circuits, an unseen circuit family, an unseen Qiskit coding style) and a QuantumKatas
regression probe, and we test the evaluator before training. Measurement counts miss every phase-only
error; a surface rule-based translator solves two of three original splits (text substitution); a
converter that \emph{executes} the Qiskit input and maps the circuit solves all 640 tasks---so the
benchmark measures source-level translation, not whether a circuit can be ported after running it.
No base model emits a valid CUDA-Q kernel zero-shot. On Nemotron 3 Nano 4B, LoRA reaches 83.9\%,
tying a 4-shot prompt (83.1\%), but generalizes worse to unseen style (48.5\% vs.\ 65.5\%) and cuts
QuantumKatas from 23.8\% to 8.9\%; on Qwen2.5-Coder 7B, LoRA beats the prompt overall (92.0\% vs.\ 82.8\%)
and on unseen style (78.5\% vs.\ 65.0\%), and on the 1.5B model as well, so whether fine-tuning pays off
depends on the base model, while the Qiskit regression appears in all three. Mixing two replay formats is
the best mitigation trade-off (84.4\% translation, Katas 21.5\%). On 48 parametric functions that must
become kernels with arguments---where executing the input cannot yield a parametric kernel---models
reach up to 83\%; on 27 real QED-C kernels, every model scores 0\%. Language models can do source-level
translation of simple parametric code, but not yet of real programs.
\end{abstract}

\section{Introduction}\label{sec:intro}
Most public quantum software is written in Qiskit, while GPU-accelerated simulation and tight
QPU--GPU integration are increasingly offered through NVIDIA CUDA-Q~\cite{cudaq}, which is also part of
the stack now used to integrate QPUs into supercomputing centres~\cite{nvidiaslurm2026}. Moving code between
the two frameworks is a practical adoption cost, and it is now being automated with language models:
NVIDIA ships an agent skill for porting Qiskit circuits to CUDA-Q~\cite{nvidiaskill}, and frontier coding
agents have ported the QED-C application benchmarks from Qiskit to CUDA-Q~\cite{brown2026187}. Two
questions remain open. Can a \emph{small open} model, which an organization can run and adapt itself,
learn this translation? And how should such a translation be evaluated, given that agents in the
QED-C port lowered the test's fidelity threshold when stuck~\cite{brown2026187} and that recent audits
found quantum code-generation benchmarks that never executed or checked the wrong
quantities~\cite{seefried2026678,coleman2026422}?

Translation between quantum frameworks is unusually well suited to measurement: the reference and the
candidate can both be executed, and their state vectors compared, so correctness needs neither a human
judge nor an LLM judge. We use this to fine-tune NVIDIA Nemotron 3 Nano 4B~\cite{nemotron3nano} with
LoRA~\cite{hu2021685} on a single L4 GPU and to compare it with a 4-shot prompt of the same model, under
an evaluation that was itself tested before any model was trained.

\paragraph{Contributions.}
\begin{enumerate}
  \item An execution-verified Qiskit$\to$CUDA-Q benchmark (640 tasks in three generalization splits;
        state-vector verifier with fidelity $\ge 0.999$; QuantumKatas regression
        probe~\cite{cruzbenito2026210}) whose evaluator was tested before training: phase-error
        negative control, fake models, surface rules vs.\ a circuit-level converter that executes the
        input~\cite{guo2025967}, and a Katas audit.
  \item A case study of fine-tuning vs.\ prompting on Nemotron 3 Nano 4B, with LoRA replicated on
        Qwen2.5-Coder 1.5B and 7B~\cite{qwen25coder}: when each wins, the Qiskit regression and its syntax-interference
        mechanism, and mitigation as a trade-off (mixed replay best among the variants we tried).
  \item Scope and transfer checks under the same verifier: a converter-hard split of 48 parametric
        functions verified at three argument settings (converter 0\%, models up to 83\%), real QED-C
        code~\cite{lubinski2023} (models 0\%, converter 100\%), prompt vocabulary sensitivity, adapter+examples, pass@5, seeds, GEPA,
        and a short GRPO pilot~\cite{agrawal2025457,shao2024300}.
\end{enumerate}

\section{Related Work}\label{sec:related}
\paragraph{Translating between quantum frameworks.} Brown et al.~\cite{brown2026187} used frontier
coding agents (Opus, GPT and Gemini) to port thirteen QED-C benchmark methods from Qiskit to CUDA-Q,
validated by noiseless tests with a fidelity floor; the agents produced working ports but took
shortcuts when stuck, including lowering a test's threshold. NVIDIA's agent skill~\cite{nvidiaskill}
encodes the gate mapping and the bit-order and precision conventions for the same pair. Q-GEAR~\cite{guo2025967}
converts Qiskit circuits to CUDA-Q kernels by rules, for GPU simulation. Li et al.~\cite{li2026841} study
LLM translation from Qiskit to PennyLane and Q\# with execution-based checks, and find that one example
in the prompt adds 8--18 points; Siavash and Moin~\cite{siavash2025480} propose LLMs as transpilers between
Qiskit and Cirq. Related classical-to-quantum and Qiskit migration studies exist~\cite{qbridge2026,suarez2026}, but
none of these fine-tunes a small open model for Qiskit$\to$CUDA-Q or measures what it costs the model
in its source framework.

\paragraph{Quantum code generation and its evaluation.} Benchmarks for quantum code generation include
Qiskit HumanEval~\cite{vishwakarma2024712}, QuanBench~\cite{guo2025779} and QuanBench+~\cite{slim2026570},
QCoder~\cite{mikuriya2025101}, QCircuitBench~\cite{yang2024961} and QHackBench~\cite{basit2025008}. They
differ in what counts as correct: unit tests, process fidelity, or distances between output
distributions. Distribution-based acceptance cannot see relative phase, which our negative control
quantifies; QEncodeBench~\cite{che2026319} likewise checks equivalence up to global phase and validates
its own verifier. Audits found benchmarks whose checks were wrong or never ran~\cite{seefried2026678}, and
harness bugs that looked like model failures~\cite{coleman2026422}; we follow their recommendation to test
the evaluator itself. Formal equivalence checking~\cite{burgholzer2020420} is an alternative to a fidelity
threshold.

\paragraph{Adapting models to a framework.} Qiskit Code Assistant~\cite{dupuis2024495} specializes a model
for Qiskit, and later work post-trains it with verifiable rewards from execution~\cite{dupuis2025907}.
PennyCoder~\cite{basit2025562} applies LoRA to a small model for PennyLane. Whether fine-tuning beats
in-context learning depends on model size and language resources~\cite{giagnorio2025085}, and in
QCircuitBench fine-tuning did not always beat few-shot prompting~\cite{yang2024961}. LoRA forgets less than
full fine-tuning but does not avoid forgetting~\cite{biderman2024673}; the cost of specialization is often
called the alignment tax~\cite{lin2023256}, and replay is a standard mitigation~\cite{riemer2025337}.
Reinforcement learning with verifiable rewards~\cite{lambert2024124,shao2024300} has been argued to
generalize better and forget less than supervised fine-tuning~\cite{chu2025161,shenfeld2025259}, and
prompt optimization such as GEPA~\cite{agrawal2025457} can rival it; we test both
(\cref{sec:alternatives}). Open models for quantum hardware tasks are appearing in the same ecosystem,
for example a vision-language model for calibration~\cite{qcaleval2026}, but none for writing or
translating CUDA-Q code.
Our error taxonomy follows the study of bugs introduced by LLM code translation of Pan et al.~\cite{pan2023109}.

\section{Task, Data and Evaluator}\label{sec:method}
This section defines the translation task, the corpus and splits, what the verifier checks, and the
tests we ran on the verifier before any model was trained. The design goal is that a reported gain
cannot come from an untested harness or from a test set that only measures surface rewrite.

We generate 1{,}946 Qiskit/CUDA-Q pairs from six circuit families (\texttt{generate/}). Each pair is
executed on both sides and kept only if state fidelity is $\ge 0.999$. Gate set:
\texttt{h}, \texttt{x}, \texttt{y}, \texttt{z}, \texttt{s}, \texttt{t}, \texttt{rx}, \texttt{ry},
\texttt{rz}, \texttt{cx}, \texttt{cz}, \texttt{swap}, plus \texttt{cp}, \texttt{cry} and \texttt{mcx}
in the QFT, W and Grover families.

\begin{table}[t]
\centering\small
\caption{Data splits. Training uses the canonical Qiskit style; each test set asks a different
generalization question.}\label{tab:splits}
\begin{tabular}{@{}lrp{0.34\linewidth}p{0.26\linewidth}@{}}
\toprule
Split & Pairs & Content & Question\\
\midrule
Train / validation & 1,355 / 151 & random, GHZ, W, QFT, Grover; 2--5 qubits & \\
Size test & 240 & same families, 6--7 qubits & generalization in size?\\
Structure test & 200 & hardware-efficient ansatz, never seen & generalization in structure?\\
Style test & 200 & test circuits rewritten with common Qiskit idioms & generalization in coding style?\\
\bottomrule
\end{tabular}
\end{table}

The style test rewrites only the Qiskit side (\cref{tab:idioms}). Gates, sizes and the CUDA-Q
reference stay the same; every rewrite was executed and matches the original state. Training data
stays in the canonical style on purpose, so the style split measures whether the model translates
semantics rather than a fixed surface form.

\begin{table}[t]
\centering\small
\caption{Qiskit idioms in the style test. Each circuit receives 2--4 idioms; QFT and Grover use only
the text-level ones so loop bodies do not explode.}\label{tab:idioms}
\begin{tabular}{@{}lp{0.72\linewidth}@{}}
\toprule
Idiom & What changes on the Qiskit side\\
\midrule
\texttt{nome} & circuit variable is not named \texttt{qc}\\
\texttt{registro} & \texttt{QuantumRegister} and indices \texttt{qr[i]}\\
\texttt{difusao} & multi-target one-qubit calls, e.g.\ \texttt{h([0, 2])} / \texttt{h(range(n))}\\
\texttt{append} & gate objects, e.g.\ \texttt{append(RYGate($\theta$), [q])}\\
\texttt{funcao} & body wrapped in a helper that takes the circuit\\
\texttt{barreira} & \texttt{barrier()} in the middle\\
\texttt{medida} & \texttt{measure\_all()} at the end (prompt still asks for a kernel without measurement)\\
\texttt{numpy} & \texttt{math.pi} rewritten as \texttt{np.pi}\\
\bottomrule
\end{tabular}
\end{table}

\paragraph{What is checked.} Explicitly, and nothing else:
\begin{enumerate}
\item the answer contains Python code (the first fenced block that mentions \texttt{cudaq}, otherwise
      the first block); an answer cut at the 1024-token limit fails as truncated, even if the
      cut-off text would parse; a final block left unclosed by an answer that ended normally is used;
\item the code defines a CUDA-Q kernel that takes no arguments;
\item the kernel compiles and runs;
\item its state has fidelity $|\langle\psi_a|\psi_b\rangle|^2 \ge 0.999$ with the Qiskit reference
      (this ignores \emph{global} phase; relative phase still changes the state and is caught).
\end{enumerate}
We report valid-output rate and state equivalence in separate columns. Environment failures are
counted apart and retried once. Each candidate runs in its own subprocess (\texttt{python -I},
temporary directory) with no access to the reference module~\cite{seefried2026678}.

\paragraph{Sanitization before \texttt{get\_state}.}
The verifier strips \texttt{mz} calls and drops top-level non-definitions (\texttt{print},
\texttt{cudaq.sample}). This is an evaluator contract, not a claim that measurement is irrelevant:
with \texttt{mz} left in, \texttt{cudaq.get\_state} collapses and a correct candidate fails at random
(see calibration). The prompt asks for a kernel without measurement; stripping \texttt{mz} avoids
punishing otherwise-correct translations for a harness artefact.

\subsection{Calibration (CUDA-Q 0.16.0, \texttt{qpp-cpu}, Qiskit 2.6.1)}\label{sec:calibration}

\begin{itemize}
\item
  \textbf{State vectors use the same qubit order} in both libraries (minimum fidelity 1.0 over 60
  random circuits with all 12 gates, up to 7 qubits). We compare vectors directly, without bit
  reversal.
\item
  \textbf{\texttt{sample} bitstrings use the opposite order} (\texttt{X} on qubit 0 gives
  \texttt{100} in CUDA-Q and \texttt{001} in Qiskit). Comparing counts naively would fail correct
  translations; our verifier therefore never scores on bitstring histograms.
\item
  \textbf{\texttt{mz} inside a kernel collapses the state returned by \texttt{get\_state}.} A
  4-qubit GHZ with \texttt{mz} returned $|0000\rangle$ or $|1111\rangle$, never the superposition.
\end{itemize}

\subsection{Testing the evaluator}\label{sec:eval-tests}

\begin{itemize}
\item
  \textbf{Negative control:} we corrupted 600 correct translations (final phase, flipped angle,
  deleted gate, swapped control). Of the 503 mutants that actually change the state, fidelity
  rejected 500; the other 3 had fidelity up to 0.9997. \textbf{Measurement counts caught 0 of the
  150 phase-only errors}, because probabilities on the computational basis do not see relative phase.
\item
  \textbf{Fake models:} a perfect model scores 100\%, garbage 0\%, and ``valid but wrong'' scores
  100\% valid output with 0.2\% equivalence---confirming that the two columns can separate.
\item
  \textbf{Rule-based baseline:} a ${\sim}70$-line translator scored 100\% on the size and family
  tests, which motivated the style test. That baseline is intentionally written for the canonical
  style; it is a diagnostic of the splits, not a production porting tool.
\item
  \textbf{Circuit-level converter:} a ${\sim}50$-line converter that \emph{executes} the Qiskit
  input, takes the resulting \texttt{QuantumCircuit} and maps its instructions to CUDA-Q, in the
  spirit of Q-GEAR~\cite{guo2025967}, scores 100\% on all 640 tasks, including the unseen style,
  because Qiskit resolves every idiom before conversion (\texttt{eval/linha\_base\_circuito.py}). Its
  kernels are flat gate lists: the loops, helper functions and parameters of the source are not
  preserved. \Cref{sec:discussion} discusses what this implies for the scope of the benchmark.
\end{itemize}

\subsection{Regression probe}\label{sec:regression-probe}

The Qiskit QuantumKatas~\cite{cruzbenito2026210,qiskitkatasrepo} (350 tasks) were audited first:
every canonical solution was run three times and every empty skeleton once. One test
(\texttt{Teleportation/1.7}) passes the empty skeleton, because the skeleton already returns
\texttt{True} and the test only checks the return value. It was excluded, leaving 349 tasks.

\section{Training and Generation}\label{sec:setup}
LoRA on \texttt{nvidia/NVIDIA-Nemotron-3-Nano-4B-BF16} (hybrid Mamba-2 + Transformer), in BF16, with Hugging Face TRL 0.22.2, PEFT and transformers 4.56.2.

\begin{table}[t]
\centering\small
\caption{LoRA configuration and the reason for each choice.}\label{tab:lora}
\begin{tabular}{@{}p{0.2\linewidth}p{0.33\linewidth}p{0.4\linewidth}@{}}
\toprule
Setting & Value & Why\\
\midrule
Rank / alpha & 32 / 64 & enough capacity for a narrow task\\
Target modules & attention (q, k, v, o) and MLP (up, down) & PEFT rejects LoRA on the Mamba \texttt{out\_proj}/\texttt{conv1d} of this model, and vLLM does not map LoRA onto the fused Mamba \texttt{in\_proj}\\
Effective batch & 16 & about 170 optimizer steps over 2 epochs\\
Learning rate & $10^{-4}$, 5\% warmup, cosine & common range for LoRA; not tuned\\
Loss & completion only & do not learn to repeat the prompt\\
Checkpoint & lowest validation loss (step 160 of 170) & selected without looking at the test sets\\
Reasoning mode & \texttt{enable\_thinking=False} everywhere & the model reasons by default\\
\bottomrule
\end{tabular}
\end{table}

\textbf{Training data uses the chat format.} With plain-text prompt/completion fields, TRL 0.22.2 tokenizes without the chat template, so training and inference would see different tokens. \texttt{train/confere\_tokens.py} runs TRL\textquotesingle s own preprocessing on CPU and checks, for every example, that the training prefix is token-identical to what generation feeds the model, that only the answer is in the loss, and that the end-of-turn token is learned.

Generation runs on vLLM 0.24 (the Red Hat AI build) on OpenShift AI, one L4, with LoRA adapters loaded at runtime (\texttt{k8s/vllm-nemotron.yaml}). The client tokenizes each prompt with the verified template and sends token ids to \texttt{/v1/completions}, so the model receives exactly the tokens checked in training. Greedy decoding, \texttt{max\_tokens} 1024, stop on \texttt{\textless{}\textbar{}im\_end\textbar{}\textgreater{}} (id 11) and \texttt{\textless{}/s\textgreater{}} (id 2).

Two findings shaped this setup:

\begin{itemize}
\tightlist
\item
  \textbf{The Hugging Face \texttt{generate} path of this model has no working cache.} The hybrid cache is created but never reaches the forward pass, and the cache class has further defects, so every token recomputes the full sequence (about 60 s per task). vLLM implements Nemotron-H natively.
\item
  \textbf{Greedy decoding is not reproducible under dynamic batching.} Running the same 40 tasks twice with 16 concurrent requests gave identical outputs for only 23 of them; with one request at a time, for 40 of 40. All evaluation runs therefore use one request at a time (about 10 s per task). With high concurrency, part of the before/after difference would be server noise.
\end{itemize}

\section{Results}\label{sec:results}
Same 640 translation tasks and 349 QuantumKatas tasks for every model, same inference engine and settings. 95\% Wilson intervals in brackets; p-values from an exact McNemar test on the tasks whose outcome changed.

\begin{figure}[t]
  \centering
  \includegraphics[width=\linewidth]{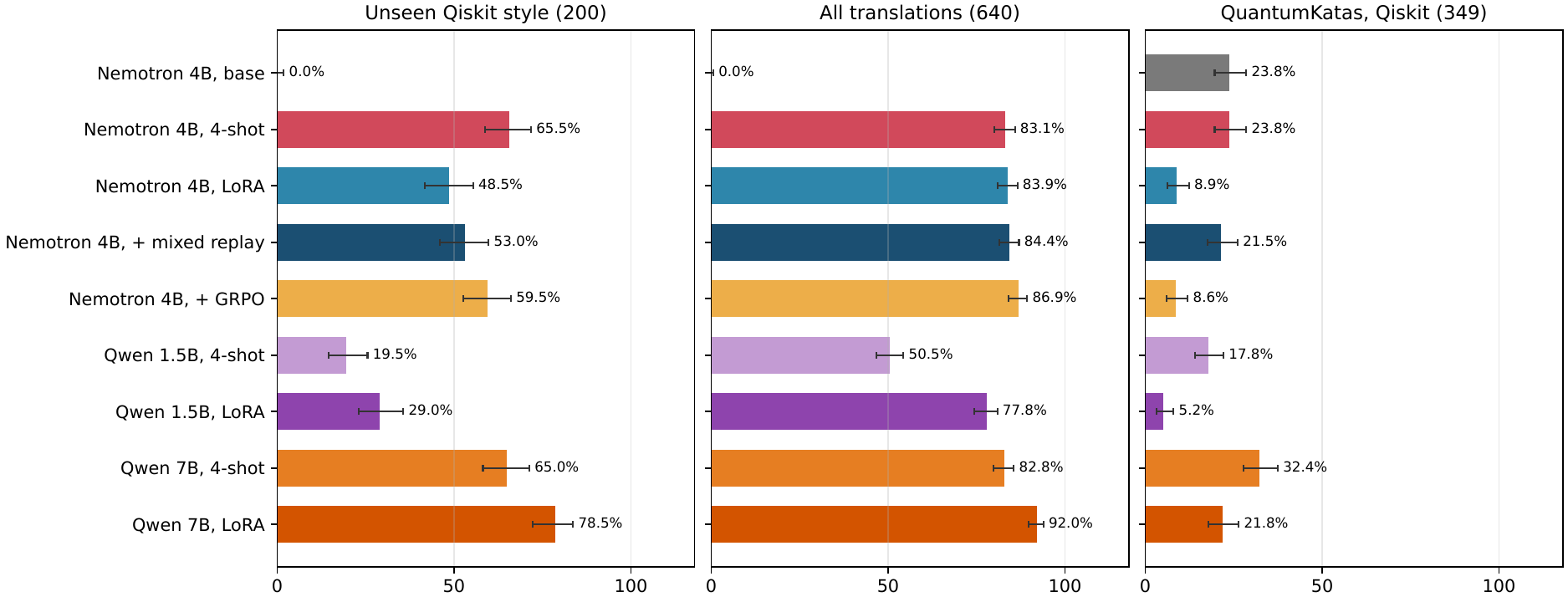}
  \caption{State equivalence on the unseen coding style and on all translations, and the QuantumKatas
regression probe (whiskers: 95\% Wilson intervals). On Nemotron 4B, LoRA ties the 4-shot prompt overall
but loses on unseen style and on the Katas; mixed replay nearly restores the Katas. On Qwen2.5-Coder
1.5B and 7B, LoRA beats the prompt, including on unseen style for the 7B, while the Qiskit regression
appears in every fine-tuned model. Few-shot Katas scores are those of the unchanged base model.}
  \label{fig:summary}
\end{figure}

\begin{table}[t]
\centering\small
\caption{Main results on the first (greedy) sample. Valid output $=$ compiles and runs as a
no-argument CUDA-Q kernel; equivalence $=$ fidelity $\ge 0.999$ against the Qiskit reference.
95\% Wilson intervals in brackets. ``Surface rules'' is a ${\sim}70$-line script for the canonical
style; ``Circuit converter'' executes the Qiskit input and maps the resulting circuit (100\% on all
splits, including style).}\label{tab:main}
\resizebox{\linewidth}{!}{%
\begin{tabular}{@{}lccccc@{}}
\toprule
& Surface rules & Circuit conv. & Base & Base + 4-shot & Fine-tuned (LoRA)\\
\midrule
\multicolumn{6}{l}{\emph{Valid output}}\\
\quad All (640)
  & 68.8\% & 100\% & 0\% & 91.4\% [89.0--93.3] & 84.2\% [81.2--86.8]\\
\midrule
\multicolumn{6}{l}{\emph{State equivalence}}\\
\quad Larger circuits (240)
  & 100\% & 100\% & 0\% & 89.6\% & \textbf{100\%} [98.4--100]\\
\quad Unseen family (200)
  & 100\% & 100\% & 0\% & 93.0\% & \textbf{100\%} [98.1--100]\\
\quad Unseen style (200)
  & 0\% & \textbf{100\%} & 0\% & 65.5\% [58.7--71.7] & 48.5\% [41.7--55.4]\\
\quad All (640)
  & 68.8\% & \textbf{100\%} & 0\% & 83.1\% [80.0--85.8] & 83.9\% [80.9--86.5]\\
\midrule
QuantumKatas (349)
  & --- & --- & \textbf{23.8\%} [19.6--28.5] & 23.8\% (unchanged) & 8.9\% [6.3--12.3]\\
\bottomrule
\end{tabular}}
\end{table}

What this shows:

\begin{itemize}
\item
  \textbf{The base model does not know CUDA-Q.} None of its 640 answers was a valid kernel. It wrote
  functions that contained only comments, or confused CUDA-Q with CUDA C
  (\texttt{\_\_global\_\_ void kernel()}).
\item
  \textbf{Valid output and equivalence tell different stories for the prompt.} The 4-shot model
  produces a runnable kernel on 91.4\% of tasks but is equivalent on 83.1\%: many failures are
  valid-looking code with the wrong state or invented gate names. LoRA is tighter (84.2\% valid,
  83.9\% equivalent): when the adapter emits a kernel, it is almost always the right one.
\item
  \textbf{On the training style, the fine-tuned model matches surface rules} (100\% on larger
  circuits and on the unseen family). Those test sets are solvable by text substitution. The
  circuit-level converter scores 100\% on every split, including style, by executing the input first
  (\cref{sec:discussion}).
\item
  \textbf{On an unseen coding style, the model generalizes partially (48.5\%), where surface rules
  score 0\% and the converter scores 100\%.} The hardest idiom (9.8\%) is
  \texttt{qc.append(RYGate($\theta$),~[q])}, for which the model invents gate names such as
  \texttt{r\_y}, or copies Qiskit structure into CUDA-Q.
\item
  \textbf{The cost in Qiskit is real: 23.8\% $\to$ 8.9\% on the QuantumKatas} (62 tasks lost, 10
  gained; $p = 3\cdot 10^{-10}$). The model did not forget Qiskit. It keeps the Qiskit function
  skeleton but writes CUDA-Q-style gate calls inside it (\texttt{r1.ctrl(q0, q1)} instead of
  \texttt{qc.cp(...)}): interference between the two syntaxes.
\item
  \textbf{A 4-example prompt matches the fine-tuned model overall} (83.1\% vs.\ 83.9\%,
  $p = 0.69$). LoRA is better on the training style (100\% vs.\ 90--93\%), where a rule-based script
  also works. The prompt is better on the unseen style (65.5\% vs.\ 48.5\%, $p = 2\cdot 10^{-5}$),
  because the untouched base model keeps its broader Qiskit knowledge. It has no Qiskit regression,
  but adds about 1{,}370 tokens to every request (about $5\times$ the zero-shot prompt: 1{,}712 vs.\
  342 tokens on average). The four examples come from the training set and are fixed for all tasks.
\end{itemize}

Training took 29 minutes on one L4 (peak 15.4\,GB). The main study (four trainings, six evaluation
runs and the tests) used about 17 GPU-hours; the checks in \cref{sec:alternatives} were run afterwards
on the same hardware.

\subsection{Sensitivity of the prompt to its examples}\label{sec:fewshot-sensitivity}
The 4-shot result above uses one fixed set of examples. We drew two more sets with the same composition
(one pair per family, same sizes, taken from the training set) and reran the base model
(\cref{tab:fewshot-sets}). One set matched the original; the other fell from 83.1\% to 48.3\%. The cause is
vocabulary coverage: the failing set contains no \texttt{rx} or \texttt{ry}, and without them the base
model scores 0\% on the W family (controlled $R_y$ rotations) and 35\% on the hardware-efficient ansatz
(layers of $R_y$), against 77--83\% and 92--93\% with the other sets; its failures become Qiskit gate names
(132 of 331). No set shows \texttt{ry.ctrl}, but with \texttt{ry} the model generalizes to the controlled form.
To test this explanation directly, we kept set 2 and replaced only its random-circuit example with one from
the training set that uses all twelve base gates (\cref{tab:fewshot-sets})---a confirmatory edit chosen
after seeing the failure, not a pre-registered selection policy. Equivalence rose from 48.3\% to
71.9\% (W family 0\%$\to$57\%, hardware-efficient ansatz 35\%$\to$73\%), so missing vocabulary explains most of
the drop, though not all of it: the other examples of the set also matter. The prompt ties the adapter only
when its examples cover the gate vocabulary of the inputs; the adapter, trained on all gates, does not have
this failure mode.

\begin{table}[t]
\centering\small
\caption{4-shot prompt with three example sets of the same composition (state equivalence).}\label{tab:fewshot-sets}
\begin{tabular}{@{}lcccc@{}}
\toprule
& All tasks (640) & Unseen style (200) & W family & HEA family\\
\midrule
Set 0 (original) & 83.1\% & 65.5\% & 77\% & 93\%\\
Set 1 & 81.9\% & 63.5\% & 83\% & 92\%\\
Set 2 (no \texttt{rx}/\texttt{ry}) & 48.3\% & 38.0\% & 0\% & 35\%\\
Set 2, random example covering all gates & 71.9\% & 58.5\% & 57\% & 73\%\\
LoRA (reference) & 83.9\% & 48.5\% & & \\
\bottomrule
\end{tabular}
\end{table}

\subsection{Error taxonomy}\label{sec:taxonomy}
To see \emph{how} each model fails, we classified every failed first sample by static analysis of the
generated code (it is not executed again), with categories adapted from Pan et al.~\cite{pan2023109}
(\texttt{eval/taxonomia.py}; \cref{tab:taxonomy}). The base model does not know CUDA-Q: it writes no kernel
or CUDA C. The 4-shot prompt knows the structure but not the vocabulary: it invents names such as
\texttt{r\_y} and uses Qiskit gate names such as \texttt{cx}. The fine-tuned models know the vocabulary but
let the input idiom leak: the LoRA model writes Qiskit API calls and copies \texttt{barrier()} or
\texttt{measure\_all()} from the unseen style; the style-augmented model (A) turns
\texttt{append(XGate(), \dots)} into class names such as \texttt{X}; replay in the Katas' function format (B)
copies Qiskit constructs and API calls (75 of 127 failures), and only 8 failures use raw integer indices.
Wrong-state failures (the code runs but prepares another state) are rare for every model.
\Cref{tab:taxonomy-variants} in the appendix extends the taxonomy to the later variants.
Later variants (mixed replay, GRPO, Qwen 1.5B) are summarized in
\cref{sec:alternatives,sec:mitigation}; we keep \cref{tab:taxonomy} focused on the main contrast.

\begin{table}[t]
\centering\small
\caption{Failure categories of the first sample, by model (640 translation tasks).}\label{tab:taxonomy}
\resizebox{\linewidth}{!}{%
\begin{tabular}{@{}lrrrrrr@{}}
\toprule
Category & Base & 4-shot & LoRA & Replay (script) & Styles (A) & Replay (function, B)\\
\midrule
No CUDA-Q kernel & 358 & -- & 3 & 2 & -- & --\\
CUDA C instead of CUDA-Q & 193 & -- & -- & -- & -- & --\\
Truncated & 89 & 9 & 1 & -- & 1 & 2\\
Qiskit API in kernel (\texttt{qc.}) & -- & -- & 42 & -- & -- & 31\\
Qiskit construct copied (\texttt{barrier}, \texttt{measure\_all}) & -- & 13 & 19 & 38 & -- & 44\\
Qiskit gate class (\texttt{X}, \texttt{CRY}) & -- & -- & 14 & -- & 39 & 1\\
Qiskit gate name (\texttt{cx}, \texttt{cp}) & -- & 24 & -- & 2 & 1 & 1\\
Invented gate (\texttt{r\_y}) & -- & 29 & 12 & 13 & 20 & 19\\
Raw integer indices & -- & -- & -- & -- & 10 & 8\\
Other syntax error & -- & -- & 4 & 10 & 7 & 2\\
Other compile error & -- & 1 & 6 & 18 & 13 & 16\\
Runs, missing operations & -- & 12 & -- & 3 & 2 & 2\\
Runs, extra operations & -- & 7 & -- & 3 & -- & 1\\
Runs, wrong qubit/order & -- & 7 & 1 & -- & -- & --\\
Runs, other wrong state & -- & 6 & 1 & 7 & 1 & --\\
\midrule
Failures (of 640) & 640 & 108 & 103 & 96 & 94 & 127\\
\bottomrule
\end{tabular}}
\end{table}

\section{Robustness Checks and Alternatives}\label{sec:alternatives}
All runs below use the same 640 translation tasks, 349 Katas, verifier and greedy decoding
(\cref{tab:alternatives}).

\begin{table}[t]
\centering\small
\caption{Robustness checks and alternatives (state equivalence, first sample). Held-out idioms are the
\texttt{append} and helper-function rewrites of the unseen style, which no training set contains. LoRA s0 is
the adapter of \cref{sec:results}; GRPO continues it with reinforcement learning on the style-augmented prompts
of variant A (\cref{sec:mitigation}).}\label{tab:alternatives}
\resizebox{\linewidth}{!}{%
\begin{tabular}{@{}lcccc@{}}
\toprule
& All tasks (640) & Unseen style (200) & Held-out idioms (120) & QuantumKatas (349)\\
\midrule
Nemotron 4B, LoRA seed 0 / 1 / 2 & 83.9 / 82.5 / 80.2\% & 48.5 / 45.0 / 37.0\% & 27.5 / 29.2 / 15.8\% & 8.9 / 6.9 / 7.4\%\\
Nemotron 4B, LoRA + style variety (A) & 85.3\% & 53.0\% & 21.7\% & 7.4\%\\
Nemotron 4B, LoRA s0 + GRPO (200 steps) & \textbf{86.9\%} & 59.5\% & 35.8\% & 8.6\%\\
Nemotron 4B, LoRA s0 + 1-shot / 2-shot prompt & 74.8 / 78.4\% & 40.0 / 46.5\% & 31.7 / 43.3\% & ---\\
Nemotron 4B, base, 4-shot & 83.1\% & \textbf{65.5\%} & \textbf{59.2\%} & 23.8\%\\
Qwen2.5-Coder 1.5B, base, 4-shot & 50.5\% & 19.5\% & 19.2\% & 17.8\%\\
Qwen2.5-Coder 1.5B, LoRA & 77.8\% & 29.0\% & 28.3\% & 5.2\%\\
Qwen2.5-Coder 7B, base, zero-shot & 0\% & 0\% & 0\% & \textbf{32.4\%}\\
Qwen2.5-Coder 7B, base, 4-shot & 82.8\% & 65.0\% & 54.2\% & 32.4\%\\
Qwen2.5-Coder 7B, LoRA & \textbf{92.0\%} & \textbf{78.5\%} & \textbf{79.2\%} & 21.8\%\\
\bottomrule
\end{tabular}}
\end{table}

\paragraph{Training seeds.} Two more LoRA runs with the same data and configuration and different seeds
(validation loss 0.0057 at the same checkpoint for all three) give 82.2\% $\pm$ 1.9 points overall but
43.5\% $\pm$ 5.9 points on the unseen style. Results on the training style are stable; generalization to
new idioms depends on the seed, and the gap to the 4-shot prompt on the unseen style (65.5\%) is larger than
that spread.

\paragraph{Adapter plus examples.} Adding one or two examples to the prompt of the seed-0 adapter does not
combine the strengths of both. With two examples, held-out idioms improve (27.5\%$\to$43.3\%; 26 tasks gained,
7 lost, $p=0.001$), but the adapter loses its exactness on the training style (size 96.7\%, family 88.5\%) and
the total falls to 78.4\% (30 gained, 65 lost, $p<10^{-3}$); one example is worse on both. The adapter was
trained without examples in the prompt, and adding them moves it away from its training distribution.

\paragraph{More samples (pass@5).} With five independent samples (first greedy, four at temperature 0.7) on
the 120 held-out-idiom tasks, the adapter goes from 28.3\% to 37.5\% and the 4-shot prompt from 58.3\% to
78.3\%, so the gap widens rather than closes. (The greedy first samples of these runs differ from the main runs
in one task each, after the server was restarted.)

\paragraph{A second model with fine-tuning.} To see whether the comparison depends on the model, we applied
the same data and LoRA configuration to Qwen2.5-Coder 1.5B-Instruct~\cite{qwen25coder} (Apache-2.0; 18 minutes
on the L4), served with vLLM 0.33 on the same GPU. On this smaller model fine-tuning clearly beats the prompt
overall (77.8\% vs.\ 50.5\%) and reaches 100\% on the training style, while both stay low on the unseen style
(29.0\% vs.\ 19.5\%); zero-shot it produced no valid kernel on a sample of 40 tasks. The Qiskit regression
appears again: 17.8\%$\to$5.2\% on the Katas. So whether fine-tuning ties or beats a prompt depends on how well
the base model learns from examples in context, consistent with~\cite{giagnorio2025085}, but the cost in the
source framework appears in both models.

\paragraph{A larger code model.} Qwen2.5-Coder 7B-Instruct~\cite{qwen25coder}, almost twice the size of
Nemotron, is no better zero-shot: none of its 640 answers is a valid kernel (it writes explanations and code
for cuQuantum or Cirq, 606 of them with syntax errors). With the same four examples it reaches 82.8\%, the
same as Nemotron (83.1\%), with 65.0\% on the unseen style. With the same data and LoRA configuration (batch 1
with 16 accumulation steps, 36 minutes and 19\,GB on the L4; served with vLLM 0.33, one request at a time),
it reaches 92.0\%, with 100\% on larger circuits and 78.5\% on the unseen style, and beats its own prompt both
overall (83 tasks gained, 24 lost, $p<10^{-8}$) and on the unseen style (43 gained, 16 lost, $p<10^{-3}$);
held-out idioms reach 79.2\%, the best of any model. The Qiskit regression is smaller but still clear:
32.4\%$\to$21.8\% on the Katas (10 gained, 47 lost, $p<10^{-6}$). The 4-shot run was repeated with one request
at a time and gave the same aggregate numbers as an earlier run with 8 concurrent requests.

\paragraph{Prompt optimization (GEPA).} GEPA~\cite{agrawal2025457} rewrites a system prompt from execution
feedback, here with Gemma as the reflection model, 300 metric calls and a graded score on a validation subset.
Starting from zero-shot it never left a score of zero: without a single valid kernel there is no signal to
reflect on. Starting from the weak example set of \cref{sec:fewshot-sensitivity}, it raised the graded
validation score from 0.73 to 0.89 and wrote an explicit gate-mapping prompt, but on the test sets it scored
43.6\% against 44.8\% for its starting prompt (unseen style 36.5\% vs.\ 31.5\%). The optimized prompt fit the
validation tasks without transferring.

\paragraph{Reinforcement learning with the verifier as reward (GRPO pilot).} We continued the seed-0
adapter with GRPO~\cite{shao2024300} (TRL, rollouts from vLLM on the same L4, 4 samples per prompt,
learning rate $10^{-5}$, no KL penalty) for 200 steps (2.5 hours), on the style-augmented prompts of
variant A, which exclude the two held-out idioms. The reward is the verifier itself, graded: 1 for an
equivalent kernel, $0.1 + 0.4F$ for a kernel that runs with fidelity $F$, 0 otherwise. The mean
reward rose from 0.86 to 0.92. In this short pilot, GRPO is the first weight update that improves
idioms it never saw: 35.8\% on the held-out idioms against 21.7\% for SFT on the same prompts (A) and
27.5\% for its starting point, with 24 tasks gained and 2 lost on the unseen style ($p<10^{-4}$), and
no further loss on the Katas (8.6\%). It also changes how the model fails: Qiskit API calls and copied
constructs disappear, but Qiskit gate class names (44) and kernels that run with a missing gate (11)
appear, consistent with a reward that pays partial credit for code that runs. It remains below the
4-shot prompt on new idioms (59.2\%) and does not repair the Qiskit regression. We treat these numbers
as a pilot, not as a tuned RL result.

\paragraph{Real-world code (QED-C).} All tasks so far are generated. As an out-of-distribution check we took
the Qiskit kernels of six QED-C application benchmarks~\cite{lubinski2023} (Bernstein--Vazirani,
Deutsch--Jozsa, phase estimation, QFT in two variants, hidden shift and Grover; Apache-2.0), removed only the
drawing helper, and appended a call that builds the circuit at 3--7 qubits, for 27 tasks
(\texttt{generate/qedc.py}). The code is what a developer wrote: functions, branches on a \texttt{method}
argument, oracles built as sub-circuits and appended, module-level state. No model translated any of them
(0/27 for the 4-shot prompt, LoRA, GRPO and mixed replay); valid kernels were 30\%, 0\%, 7\% and 11\%. The
models either copy the source's structure into the kernel (variables, the \texttt{method} branch, an invented
\texttt{cudaq.circuit}), loop on a repeated gate pattern until the token limit, or guess a circuit that does
not match. The circuit-level converter, which executes the code and transpiles the result, translates all 27.
Learning the API from generated circuits does not transfer to reading real programs, which is the setting where
a language model would be needed.

\paragraph{A converter-hard split: parametric functions.} To test the setting where the converter cannot
apply, we generated 48 Qiskit functions with parameters in eight families (a loop over the number of qubits,
a scalar angle, a list of angles with layers, the QFT, a branch on a boolean argument, an integer encoded in
bits, controlled rotations from a list, and helper functions), six variants each (gates, entangler, names,
register style), and asked for a CUDA-Q kernel with the same parameters and type annotations
(\texttt{generate/converter\_hard.py}). Each task is verified at three argument settings (for example
$n=3,4,6$ with different angles); a task counts only if all three states match the Qiskit function's
($\geq 0.999$). The reference kernels pass all 48 tasks. The converter, even when allowed to execute the
function with the first argument setting, produces a kernel without parameters and fails all 48. Models do
much better (\cref{tab:hard}): the 4-shot prompt translates 48\%, the adapter 63\%, GRPO 71\%, mixed replay
83\% (21 gained and 4 lost against the prompt, $p<10^{-3}$), and the Qwen 7B adapter 79\%; the base model
without examples writes no kernel. The hardest families are the boolean branch and helper functions, where
models inline one path or drop the helper loop. Unlike QED-C, these functions are short and use the training
gate set, so the split shows that source-level translation of simple parametric code is within reach, not
that real programs are.

\begin{table}[t]
\centering\small
\caption{Converter-hard split: 48 parametric Qiskit functions translated into CUDA-Q kernels with arguments,
each verified at three argument settings. A task passes only if all three states match. The circuit-level
converter cannot produce a parametric kernel.}\label{tab:hard}
\begin{tabular}{@{}lcc@{}}
\toprule
& Valid kernel & Equivalent at all three settings\\
\midrule
Circuit-level converter (executes with the first setting) & 0\% & 0\%\\
Nemotron 4B, base, zero-shot & 0\% & 0\%\\
Nemotron 4B, base, 4-shot & 52\% & 48\%\\
Nemotron 4B, LoRA & 73\% & 63\%\\
Nemotron 4B, LoRA + GRPO & 85\% & 71\%\\
Nemotron 4B, LoRA + mixed replay & 92\% & \textbf{83\%}\\
Qwen2.5-Coder 7B, base, 4-shot & 90\% & 73\%\\
Qwen2.5-Coder 7B, LoRA & 85\% & 79\%\\
\bottomrule
\end{tabular}
\end{table}

\section{Mitigating the Qiskit Regression}\label{sec:mitigation}
Four variants of the training data, each with the same LoRA configuration, validation set and inference engine. Nothing from the Katas or the test sets enters training.

\begin{table}[t]
\centering\small
\caption{Mitigating the Qiskit regression: four variants of the training data, same LoRA configuration
and evaluation. Held-out idioms are those absent from variant A's training data.}\label{tab:mitigation}
\resizebox{\linewidth}{!}{%
\begin{tabular}{@{}lcccccc@{}}
\toprule
State equivalence & Base + 4-shot & LoRA & + replay, script & + style variety (A) & + replay, function (B) & + replay, mixed\\
\midrule
Training style (size 240 / family 200) & 89.6\% / 93.0\% & 100\% / 100\% & 100\% / 99.0\% & 100\% / 100\% & 100\% / 99.5\% & 99.6\% / 97.5\%\\
Unseen style (200) & \textbf{65.5\%} & 48.5\% & 53.0\% & 53.0\% & 37.0\% & 53.0\%\\
\quad idioms seen in A's training (80) & 75.0\% & 80.0\% & 53.8\% & \textbf{100\%} & --- & 60.0\%\\
\quad held-out idioms \texttt{append} / helper function (120) & \textbf{59.2\%} & 27.5\% & 52.5\% & 21.7\% & --- & 48.3\%\\
All translations (640) & 83.1\% & 83.9\% & 85.0\% & 85.3\% & 80.2\% & 84.4\%\\
\textbf{QuantumKatas (349)} (base 23.8\%) & 23.8\% & 8.9\% & 15.2\% & 7.4\% & 20.3\% & \textbf{21.5\%}\\
\bottomrule
\end{tabular}}
\end{table}

\textbf{Replay, script format.} 20\% of training examples in the reverse direction (CUDA-Q → Qiskit), from training circuits. The Katas recover from 8.9\% to 15.2\% (27 gained, 5 lost; p = 1·10⁻⁴) at no cost to translation, but stay below the base model (p = 4·10⁻⁵). The syntax interference disappears: Katas answers containing CUDA-Q syntax go from 35\% (LoRA) to 2\%, and answers using \texttt{qc.\textless{}gate\textgreater{}(...)} from 6\% to 70\% (base: 0\% and 78\%). What remains is a format problem: of the 41 tasks the base model solved and this model fails, only 21 contain \texttt{return\ qc} (base: 41 of 41). The reverse examples are flat scripts with no function or \texttt{return}, and the model carried that format into the Katas. Gate-logic errors also appear, such as \texttt{qc.cz} where the base model wrote \texttt{qc.cx(control,\ target)}.

\textbf{(A) Style variety.} 30\% of training pairs rewritten with 6 idioms (replacing the originals, so the training set keeps its size), holding out the \texttt{append} and helper-function idioms for the test. Seen idioms reach 100\%; held-out idioms stay at 21.7\%. Fine-tuning learns the variations it is shown and does not become more robust to new ones. Only the untouched base model with a prompt does well on idioms nobody showed it (59.2\%).

\textbf{(B) Replay, function format.} 339 Qiskit examples in the Katas\textquotesingle{} format ("complete the function and return the circuit"), with docstrings that describe the gates, built from training circuits and verified by execution. The Qiskit regression closes (20.3\% vs 23.8\% for the base model, p = 0.10), but translation of non-canonical inputs drops to 37.0\%. Most failures copy Qiskit constructs into the kernel (\texttt{barrier()}, \texttt{measure\_all()}: 44 of 127) or write Qiskit API calls (\texttt{qc.}: 31); a smaller group (8) uses raw integer indices (\texttt{rx(0.38,\ 0)}) without allocating \texttt{cudaq.qvector}, the index style of the replay data.

\textbf{Replay, mixed.} Half of each replay set (169 reverse scripts and 170 functions), so the total and the
replay share are the same as in the two single-format variants (\texttt{train/dados\_sft.py}). It is the best
trade-off we found: translation as with script replay (84.4\%, unseen style 53.0\%, held-out idioms 48.3\%)
and the Katas almost back to the base model (21.5\% vs.\ 23.8\%; 20 gained, 28 lost, $p=0.31$). The two leaks
partly cancel: Qiskit API calls and raw indices disappear from the translations, but copied \texttt{barrier()}
and \texttt{measure\_all()} remain (47 of 100 failures), and the unseen style stays well below the prompt.

\textbf{Pattern across all variants:} auxiliary training data fixes the task it imitates and leaks its format into neighbouring tasks. Mitigation is a trade-off, and it has to be measured on both sides.

\section{Discussion: When Does Fine-Tuning Pay Off?}\label{sec:discussion}
On this task a 4-shot prompt and a LoRA adapter reach the same overall equivalence (83.1\% vs.\
83.9\%, $p=0.69$), but they fail differently and cost differently. LoRA is exact on the training
style (100\%), where a short rule-based translator also works, and it needs no examples in the
prompt; the prompt generalizes better to code written in an unseen style (65.5\% vs.\ 48.5\%) and
leaves the model's Qiskit ability untouched, but adds about 1{,}370 tokens to every call. The prompt
is also fragile in a way the adapter is not: an example set of the same composition that omitted two
rotation gates dropped it to 48\% (\cref{sec:fewshot-sensitivity}).

\paragraph{Practical consequence.} Start with a prompt whose examples cover the input gate
vocabulary, and check that coverage. Whether to fine-tune is model-dependent: on Nemotron 4B a 4-shot
prompt ties LoRA overall and wins on unseen style; on Qwen2.5-Coder 1.5B and 7B, LoRA beats the prompt,
including on style for the 7B. Fine-tune when the base model learns poorly in context, when call volume
makes extra tokens expensive, or when inputs follow a predictable format, and serve the adapter only where
it helps (vLLM can host several LoRA adapters on one base model). When inputs vary, training on varied
data fixed only the idioms that were seen, and adding examples to the Nemotron adapter's prompt cost more
on the training style than it gained on new idioms.

The regression probe changes the picture for any team that fine-tunes in place. Specializing a 4B
model on one translation direction cost almost two thirds of its QuantumKatas score, and the error
taxonomy shows why: the model did not forget Qiskit; it wrote CUDA-Q syntax inside Qiskit functions.
Every auxiliary dataset we tried fixed the task it imitated and leaked its own format into the
neighbouring task, so mitigation has to be measured on both sides.

\paragraph{Scope: when is a language model needed at all?} The circuit-level converter
(\cref{sec:eval-tests}) solves every task, because every input in the benchmark is a static circuit that
can be executed at translation time; for such inputs, running the Qiskit code and converting the circuit
is exact and cheap. What the benchmark measures is whether a model learns \emph{source-level}
translation, which matters where execution is not available or not wanted: kernels with parameters or
arguments, classical control flow, partial programs, code that calls other libraries, or a port that
should keep the loops and functions a developer wrote. Our tasks exercise the vocabulary and idioms of
that setting, so the 640 tasks are a necessary test rather than a sufficient one. The converter-hard
split is the first positive evidence for the role of a language model where the converter fails by
construction: on short parametric functions, fine-tuned models translate up to 83\%. The QED-C check
shows the limit: on real benchmark code with functions, branches and sub-circuits, none of the models
produced a correct kernel. Richer control flow, partial programs and cross-library calls remain open.

\paragraph{Translation versus idiomatic porting.} State equivalence accepts any kernel that prepares the
right state, so it already accepts some idiomatic rewrites (a gate broadcast over a register instead of a
loop, a multi-controlled gate instead of its decomposition), but it cannot tell a naive line-by-line port from
a good one. More importantly, every task is state preparation without measurement, observables or
optimization, which leaves out the parts of a program where the two frameworks differ most
(\cref{tab:idiomatic}). A good port of a Qiskit program that estimates an energy with an \texttt{Estimator}
should not translate the primitive call by call but use \texttt{cudaq.observe} with a spin operator; an
inverted sub-circuit becomes \texttt{cudaq.adjoint}; a parameter loop becomes a kernel with arguments inside
the optimizer; a measurement-conditioned operation becomes ordinary control flow on \texttt{mz} inside the
kernel. Evaluating such ports requires comparing results (expectation values, distributions) rather than
states, and checking which native constructs the code uses. This is also where a language model has an
advantage that no circuit-level converter can match, and it is the natural next benchmark.

\begin{table}[t]
\centering\small
\caption{Where an idiomatic CUDA-Q port differs from a translation of the Qiskit code. None of these
constructs appears in our tasks, which are state preparation without measurement.}\label{tab:idiomatic}
\begin{tabular}{@{}p{0.45\linewidth}p{0.47\linewidth}@{}}
\toprule
Qiskit & Idiomatic CUDA-Q\\
\midrule
\texttt{Estimator} with a \texttt{SparsePauliOp} observable & \texttt{cudaq.observe(kernel, spin\_op, ...)} with \texttt{cudaq.spin}\\
\texttt{circuit.inverse()}, \texttt{compose()} & \texttt{cudaq.adjoint(kernel, ...)}, kernels calling kernels\\
\texttt{Parameter} and \texttt{assign\_parameters} in an optimization loop & kernel arguments with \texttt{cudaq.observe} inside the optimizer, or \texttt{cudaq.vqe}\\
Mid-circuit measurement with \texttt{if\_test} / \texttt{c\_if} & \texttt{mz} and an ordinary \texttt{if} inside the kernel\textsuperscript{a}\\
Sweeps over many parameter values & batched \texttt{observe}/\texttt{sample} calls and GPU or multi-QPU targets\\
\bottomrule
\end{tabular}\\[0.35em]
{\footnotesize\textsuperscript{a}In CUDA-Q~0.16, kernels with mid-circuit measurement and classical control must be executed with \texttt{cudaq.run} (returning a value); \texttt{cudaq.sample} rejects them.}
\end{table}

For evaluation, three lessons transfer to other quantum code-translation benchmarks. Compare states,
not measurement counts: counts missed every phase-only error in our negative control. Run a
rule-based baseline before training: it revealed that two of our three original test sets measured
text substitution. And keep the evaluator out of the model's reach: an isolated verifier with
declared checks cannot be weakened by the system it evaluates, which is precisely what happened with
coding agents in~\cite{brown2026187}.

The alternatives (\cref{sec:alternatives}) sharpen this picture. Prompted, a larger code model ties the
4B model; fine-tuned, both the 1.5B and the 7B Qwen models beat their own prompts, the 7B one also on unseen
style. The Nemotron tie is therefore a property of that model rather than of the task, and the balance between
fine-tuning and prompting has to be measured per model; the Qiskit regression, in contrast, appeared in all
three. More samples widen the prompt's advantage on new idioms (pass@5 78\% vs.\ 38\%). Automatic prompt optimization did not transfer from its validation tasks. The GRPO
pilot is the only weight update that improved idioms absent from training, but it still trails a
four-example prompt and leaves the Qiskit regression in place. A state-comparing verifier is what
makes such RL possible without a human or LLM judge.

\section{Limitations and Threats to Validity}\label{sec:limitations}
\begin{itemize}
\item
  \textbf{Scope.} Up to 7 qubits, noiseless simulation, fixed gate set. The main 640-task benchmark has
  no mid-circuit measurement and no kernel arguments. The converter-hard split adds short parametric
  kernels (including a boolean branch) but not rich classical control flow, library calls or partial
  programs. This remains a translation-of-API study, not a claim about large-scale algorithm ports, and it
  measures semantic translation of state preparation, not idiomatic porting of programs that measure,
  estimate observables or optimize (\cref{tab:idiomatic}).
\item
  \textbf{Synthetic data.} Training, the 640-task tests and the converter-hard split are generated; the
  style idioms are author-chosen (\cref{tab:idioms}). Real GitHub or tutorial Qiskit may be harder. The
  rules' 0\% on the style split reflects a translator written for the canonical style; a converter that
  executes the input solves every \emph{static} task in the main benchmark
  (\cref{sec:discussion}).
\item
  \textbf{Calibration and threshold.} Calibration holds for CUDA-Q 0.16.0 / \texttt{qpp-cpu} /
  Qiskit~2.6.1. The 0.999 threshold lets very small differences through (3 of 503 mutants), but
  results do not depend on it: at 0.99 or 0.9999 no model changes by more than one task
  (\texttt{eval/limiar.py}).
\item
  \textbf{Models and training.} Weight updates on Nemotron 3 Nano 4B (all variants) and one LoRA each on
  Qwen2.5-Coder 1.5B and 7B; mitigation, GEPA and GRPO only on Nemotron. One LoRA configuration, no hyperparameter search. Three seeds
  show that unseen-style results vary by several points. The validation set is from the training
  distribution and only selects the checkpoint.
\item
  \textbf{Inference.} Changing the engine changes exact outputs (3 tasks compared between transformers
  and vLLM: 1 identical, 2 diverged with the same content), so all reported runs share one engine.
  The Qwen runs used vLLM 0.33 on the workbench GPU (all Qwen comparisons on the same engine); an
  earlier Qwen 7B prompting run with 8 concurrent requests gave the same aggregate numbers as the one
  reported. After a server restart, a few greedy outputs of
  Nemotron differed in one task per run.
\item
  \textbf{Small checks.} pass@5 was measured only on the 120 held-out-idiom tasks, and the QED-C set has 27
  tasks from six benchmarks, enough to show a complete failure but not to rank models. The converter-hard
  split (48 tasks) is synthetic and uses the training gate set.
\item
  \textbf{Prompting and pilots.} The few-shot baseline uses fixed sets of four examples; we tested
  three, and one scored much lower (\cref{sec:fewshot-sensitivity}). GEPA and GRPO were run once each,
  with small budgets (300 metric calls; 200 steps).
\end{itemize}

\section{Conclusion}\label{sec:conclusion}
Qiskit-to-CUDA-Q translation can be measured without a judge by executing both programs and comparing
states. Doing so carefully changed our conclusions more than any training choice: measurement counts
would have accepted wrong phases; two of three original splits measured text substitution; and a
converter that runs the input ports every static circuit in the benchmark and in a QED-C sample.
Against that bar, fine-tuning a 4B open model learns the generated-API task (0\% to 84\%) but does not
beat a 4-shot prompt overall, generalizes worse to unseen style, and costs most of the model's Qiskit
ability; on Qwen2.5-Coder 1.5B and 7B, fine-tuning does beat the prompt (92\% on the 7B, also on unseen
style), with the same kind of cost. Whether fine-tuning pays off depends on the model; the regression does
not. Mixed replay nearly closes the regression without hurting translation. For static, executable circuits
a language model is not needed; for short parametric functions it helps (up to 83\%); for real QED-C kernels
it does not (0/27). The open problem is source-level translation of real programs---and, beyond that,
idiomatic porting of measurement, observables and optimization (\cref{tab:idiomatic}).

For practitioners: cover gate vocabulary in prompt examples; fine-tune when volume, format stability or
weak in-context learning justify it; serve the adapter only for translation; measure the cost in the
source framework; and prefer a circuit-level converter whenever the source can be executed at porting
time.

\paragraph{Future work.} A benchmark of idiomatic porting (\texttt{Estimator} to \texttt{cudaq.observe},
inverses to \texttt{cudaq.adjoint}, variational loops, mid-circuit measurement via \texttt{cudaq.run}),
evaluated on results and on the native constructs used; training and test data from real programs,
extending the converter-hard split to classical control flow, partial programs and code that calls other
libraries; mitigation and RL on the larger models; longer verifier-as-reward RL with mixed replay and
without partial credit for wrong-state kernels; mid-circuit measurement in the evaluator. We have started
the first of these: the task generator, the two-column evaluator (correct, idiomatic) and its controls for
\texttt{cudaq.observe}, \texttt{cudaq.adjoint} and \texttt{cudaq.control} are available at
\url{https://github.com/anapaulaappel/qiskit-to-cudaq-idiomatico}.

\section{Reproducibility}\label{sec:repro}
Code, the generated corpus, the test sets, the evaluator, the training and generation scripts and every
result file are available at \url{https://github.com/anapaulaappel/qiskit-to-cudaq} under the Apache License
2.0. The QuantumKatas are not redistributed (CC-BY-NC-SA-4.0); the repository explains how to obtain and
audit them. The base models are governed by their own licenses (NVIDIA Open Model License; Apache-2.0 for
Qwen2.5-Coder). The evaluator kills a candidate above 3~GB of memory: on CUDA-Q 0.16 (\texttt{qpp-cpu}), a
gate whose control is also its target, such as \texttt{x.ctrl(q[0], q[0])}, compiles and then uses unbounded
memory during simulation.

\appendix
\section{Error taxonomy of the later variants}\label{app:taxonomy}
\begin{table}[H]
\centering\small
\caption{Failure categories of the first sample (640 translation tasks) for the variants of
\cref{sec:alternatives,sec:mitigation}, with the LoRA adapter for reference. Mixed replay removes Qiskit API
calls but keeps copied constructs; GRPO trades copied constructs for Qiskit gate class names and missing
gates; examples in the adapter's prompt add wrong-state kernels (missing or extra gates); the Qwen adapters
mostly copy \texttt{barrier()} and \texttt{measure\_all()}.}\label{tab:taxonomy-variants}
\resizebox{\linewidth}{!}{%
\begin{tabular}{@{}lrrrrrr@{}}
\toprule
Category & LoRA & + mixed replay & + GRPO & + 2-shot prompt & Qwen 1.5B LoRA & Qwen 7B LoRA\\
\midrule
Qiskit construct copied (\texttt{barrier}, \texttt{measure\_all}) & 19 & 47 & -- & 36 & 110 & 34\\
Invented gate (\texttt{r\_y}) & 12 & 28 & 12 & 30 & 6 & --\\
Qiskit gate class (\texttt{X}, \texttt{CRY}) & 14 & -- & 44 & 14 & -- & --\\
Qiskit API in kernel (\texttt{qc.}) & 42 & -- & -- & 1 & -- & --\\
Qiskit gate name (\texttt{cx}, \texttt{cp}) & -- & 1 & 1 & 1 & -- & --\\
Raw integer indices & -- & 1 & 3 & -- & -- & --\\
No CUDA-Q kernel / truncated & 4 & -- & 1 & 2 & 1 & --\\
Other syntax error & 4 & 6 & 1 & 7 & -- & 11\\
Other compile error & 6 & 7 & 9 & 1 & 22 & 2\\
Runs, missing operations & -- & 6 & 11 & 29 & -- & 2\\
Runs, extra operations & -- & 4 & -- & 15 & 3 & 1\\
Runs, wrong qubit/order or other state & 2 & -- & 2 & 2 & -- & 1\\
\midrule
Failures (of 640) & 103 & 100 & 84 & 138 & 142 & 51\\
\bottomrule
\end{tabular}}
\end{table}

\bibliographystyle{plainnat}
\bibliography{refs}
\end{document}